\documentclass[aps,prl,twocolumn,superscriptaddress,showpacs,10pt]{revtex4-1}

\usepackage[T1]{fontenc}
\usepackage{color,graphicx}
\usepackage{subfig,array}
\usepackage[dvipdfmx,colorlinks=true]{hyperref}
\usepackage{amsthm,amssymb,amsmath}

\newcommand\ket[1]{\ensuremath{|#1\rangle}}

\newcommand\oprod[2]{\ensuremath{|#1\rangle\langle#2|}}

\begin{document}


\title{Complete Characterization of the Ground Space Structure of\\%
  Two-Body Frustration-Free Hamiltonians for Qubits}

\author{Zhengfeng Ji}%
\affiliation{Perimeter Institute for Theoretical Physics, Waterloo,
  Ontario, Canada}%
\affiliation{State Key Laboratory of Computer Science, Institute of
  Software, Chinese Academy of Sciences, Beijing, China}%
\author{Zhaohui Wei}%
\affiliation{Centre for Quantum Technologies, National University of
  Singapore, Singapore 117543, Singapore}%
\author{Bei Zeng}%
\affiliation{Department of Mathematics $\&$ Statistics, University of
  Guelph, Guelph, Ontario, Canada}%
\affiliation{Institute for Quantum Computing, University of Waterloo,
  Waterloo, Ontario, Canada}%

\begin{abstract}
  The problem of finding the ground state of a frustration-free
  Hamiltonian carrying only two-body interactions between qubits is
  known to be solvable in polynomial time. It is also shown recently
  that, for any such Hamiltonian, there is always a ground state that
  is a product of single- or two-qubit states. However, it remains
  unclear whether the whole ground space is of any succinct structure.
  Here, we give a complete characterization of the ground space of any
  two-body frustration-free Hamiltonian of qubits. Namely, it is a
  span of tree tensor network states of the same tree structure. This
  characterization allows us to show that the problem of determining
  the ground state degeneracy is as hard as, but no harder than, its
  classical analog.
\end{abstract}

\date{Oct. 12, 2010}


\maketitle

Quantum spin models are simplified physical models for real materials,
but are believed to capture some of their key physical properties,
which lie in the heart of modern condensed matter theory~\cite{Nac06}.
Ground states of strongly correlated spin systems is usually highly
entangled, even if the system Hamiltonian carries only local
interactions. So in general, finding the ground state of such a system
is intractable with traditional techniques, such as mean field theory.

In practical spin systems, different local terms in the Hamiltonian
might also compete with each other, a phenomenon called frustration,
which makes the system further difficult to analyze~\cite{Die04}.
However, frustration is not a necessary factor to cause ground state
entanglement. Frustration-free Hamiltonians can carry lots of
interesting physics, ranging from gapped spin chains~\cite{AKLT87} to
topological orders~\cite{KL09,LW06}.

During recently years, the active frontier of quantum information
science brings new tools to study quantum spin systems. In particular,
local Hamiltonian problems are shown to be in general very hard, i.e.,
QMA-complete~\cite{KSV02}. It is also realized that the study of
$k$-local frustration-free Hamiltonians for qubits is closely related
to the quantum $k$-satisfiability problem (Q-$k$-SAT)~\cite{Bra06},
which is the quantum analogy of the classical $k$-satisfiability
($k$-SAT), a problem that is of fundamental importance and has been
extensively studied in theoretical computer science (see,
e.g.,~\cite{Sip05}).

Spin models with two-body interaction are of the most physical
relevance, as two-body interaction, in particular of nearest neighbor
or next nearest neighbors on certain type of lattices, are the
strongest interaction terms in the real system Hamiltonian. Because
two-level systems are most common in nature, spin-$1/2$ (qubit)
systems are of particular importance.

It is realized, however, that certain ground states of a two-body
frustration-free (2BFF) Hamiltonian of qubits could be pretty trivial
with almost no entanglement at all. Algorithmically, the problem of
finding the ground state of a 2BFF Hamiltonian of qubits is known to
be solvable in polynomial time~\cite{Bra06}. It is also shown recently
that for any such Hamiltonian, there is always a ground state that is
a product of single- or two-qubit states; and if there is a genuine
entangled ground state, the ground space must be
degenerate~\cite{CCD+10}. There are also similar observations of the
ground states in random or generic
instances~\cite{LMSS10,BMR09,LLM+09,AKS09}, saying that the entire
ground space is of a trivial structure, which is almost always the
fully symmetric space, with ground space degeneracy $n+1$, where $n$
is the number of qubits~\cite{LMSS10,BMR09,BOOE10}.

The main purpose of this work is to characterize the entire ground
space in the most general setting. We improve the understanding of the
ground space of 2BFF Hamiltonians of qubits by showing that it is
always a span of tree tensor network states of the same tree
structure. In other words, these states are generated, from products
of single qubit states, by the same series of isometries (from single
qubit to two qubits). As this characterization holds for the most
general case, it implies that computing the degeneracy of 2BFF
Hamiltonian (\#Q-2-SAT) is in a complexity class called
\#P~\cite{Val79}. On the other hand, the classical analog \#2-SAT of
\#Q-2-SAT is \#P-hard, therefore \#P-complete.This answered a question
raised in~\cite{BMR09}.

{\em Two-body frustration-free Hamiltonian.---\/} Consider a system of
$n$ qubits labeled by the set $V=\{1,2,\ldots,n\}$. We will be
interested in 2BFF Hamiltonians $H=\sum H_J$ of the system. The
Hamiltonian is called two-body if each term $H_J$ acts non-trivially
only on two qubits. The index $J$ indicates the two qubits on which
$H_J$ acts. The Hamiltonian $H$ is called frustration-free if its
ground state also minimizes the energy of each term $H_J$
simultaneously. Without loss of generality, we can assume throughout
the paper that the smallest eigenvalue of each term $H_J$ is zero by
shifting the energy spectrum. In this convention, the frustration-free
Hamiltonian $H$ itself will have zero ground energy. Specifically, we
have
\begin{equation}
  \label{eq:intersection}
  \mathcal{K}(H) = \bigcap\,
  \bigl(
    \mathcal{K}(H_J)\otimes \mathcal{H}_{\bar{J}}
  \bigr),
\end{equation}
where $\mathcal{K}(H)$ is the ground space of $H$ and
$\mathcal{H}_{\bar{J}}$ is the Hilbert space of the qubits not in $J$.
From this equation, one easily sees that it is the ground space of
each term $H_J$, not the structures of excited states, that matters
for the ground space of a frustration-free Hamiltonian $H$. In other
words, it suffices to consider local terms to be projections $\Pi_J$
for our purpose.

Closely related to the analysis of 2BFF qubit Hamiltonians is the
quantum 2-SAT problem (Q-2-SAT) first considered by
Bravyi~\cite{Bra06}. Naturally generalizing classical 2-SAT, the
Q-2-SAT problem asks whether, for a given set of two-qubit projections
$\{\Pi_J\}$ of an $n$-qubit system, there is a global state
$\ket{\Psi}$ such that $\Pi_J\ket{\Psi}=0$ for all $J$. Apparently, we
answer ``yes'' to the problem if and only if the Hamiltonian
$\sum\Pi_J$ is frustration-free. It was known that Q-2-SAT is
decidable in polynomial time on a classical computer~\cite{Bra06}. The
proof of the statement actually constructs a specific $n$-qubit state
$\ket{\Psi}$ in the ground space of $\sum\Pi_J$ if there is any. Our
techniques will be similar to those used by Bravyi, but we will show a
stronger result that one can not only find one state in the ground
space, but also represent the entire ground space in terms of a span of
special states.


{\em A case study of the rank.---\/} Given a 2BFF Hamiltonian $H=\sum
H_J$, what can we say about the ground space $\mathcal{K}(H)$? First
of all, as argued previously, we only need to consider Hamiltonians of
the form $H=\sum\Pi_J$ where $\Pi_J$'s are projections onto
$\mathcal{K}(H_J)^{\perp}$. We will start our analysis by considering
the rank of the projections $\Pi_J$.

First, if there is a $\Pi_J$ of rank $3$, the only possible state for
the two qubits in $J$ is $I-\Pi_J$ of rank $1$, and this reduces to a
problem on qubits in $V\setminus J$.

If there is a $\Pi_J$ of rank $2$, the state of qubits in $J$ is
restricted to a two-dimensional subspace. Let $\ket{\psi_0}_{a,b}$ and
$\ket{\psi_1}_{a,b}$ be two orthogonal states that span the subspace,
where $a,b$ are the two qubits in $J$. One can encode qubits $a$ and
$b$ by a single qubit $d$. For this purpose, we define an isometry $U$
in the following form $U: \ket{0}_d \mapsto \ket{\psi_0}_{a,b}, \;
\ket{1}_d \mapsto \ket{\psi_1}_{a,b}$.
This procedure produces a set of constraints on $n-1$ qubits. It is
easy to verify that a state $\ket{\Psi}$ is in the ground space of the
reduced problem if and only if $U\ket{\Psi}$ is in the ground space of
the original problem~\cite{Bra06,CCD+10}.

When there is no projection of rank larger than $1$, we are dealing
with the homogeneous case~\cite{Bra06}. It turns out that the
homogeneous case is the hardest and we will discuss it two separate
sections. As we will see, the ground space of the homogeneous
Hamiltonian (more precisely, the simplified homogeneous Hamiltonian
defined later) is spanned by single-qubit product states. The above
case analysis gives an explicit representation of the ground space of
a general 2BFF qubit Hamiltonian, which is given by the following

\textbf{Main Observation} --- {\it The ground space is always a span of
tree tensor network states of the same tree structure.}

\begin{figure}[htbp]
  \centering
  \includegraphics{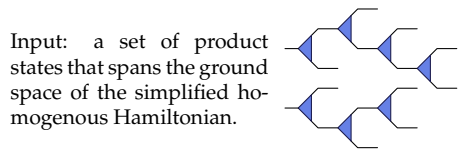}
  \caption{The general structure of the ground space}
  \label{fig:forest}
\end{figure}

We illustrate this observation in Fig.~\ref{fig:forest}, where the
ground space is viewed as a span of states generated by the isometries
(blue triangles) organized in a forest form (a collection of trees)
acting on product states (input from the left). In the language of
tensor network states~\cite{PVMC07,SDV06}, one can also represent
these states in terms of tree tensor network after combining the input
product states and the roots of trees in the forest.

{\em Homogeneous case with product constraints.---\/} Consider the
Hamiltonian $H=\sum \Pi_J$ where $\Pi_J$'s are rank-$1$ projections.
One can visualize the interactions in $H$ by a graph $G$. The graph has
$n$ vertices corresponding to the qubits and two vertices are
connected when there is a non-trivial interaction $\Pi_J$ acting on
them. We will also distinguish two types of edges in the interaction
graph. Let $\Pi=\oprod{\phi}{\phi}$ be a projection. We will use a
solid edge in the graph when $\ket{\phi}$ is entangled and a dashed
edge when $\ket{\phi}$ is a product state. Let us first focus on the
homogeneous case with product constraints only.

In this case, the interaction graph consists of dashed edges. We will
show that the ground space is a span of product of single-qubit states
(or, for simplicity, a product span). It will also be useful to know
that the states we choose are orthogonal up to a local operation $L =
\bigotimes_{j=1}^n L_j$, where $L_j$ is a non-singular local operator
acting on the $j$-th qubit. Note applying $L$ on the 2BFF Hamiltonian
$H=\sum H_J$ results in $H^L = \sum L_J^{-1} H_J L_J$, where $L_J =
\bigotimes_{j\in J} L_j$. And $H^L$, which is also 2BFF, has the same
ground state degeneracy as $H$~\cite{Bra06,CCD+10}. The relation
between the ground space of $H$ and $H^L$ is
\begin{equation}
  \label{eq:L}
  L^{-1} \mathcal{K}(H) = \mathcal{K} (H^L).
\end{equation}

\begin{figure}[htbp]
  \centering
  \begin{tabular}{m{.35\linewidth}m{.45\linewidth}}
    \subfloat[][An illustration of a dashed interaction graph]{\includegraphics{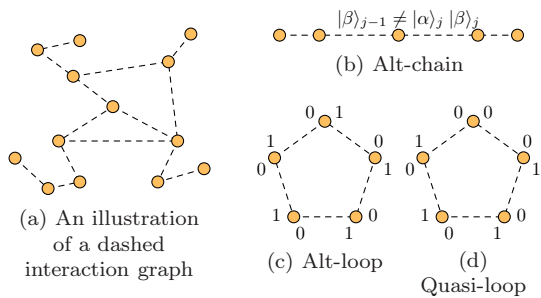}\label{fig:toy_dashed}} &
    \begin{tabular}{c}
      \subfloat[][Alt-chain]{\includegraphics{fig-7.eps}\label{fig:achain}}\\
      \begin{tabular}{cc}
        \subfloat[][Alt-loop]{\includegraphics{fig-5.eps}\label{fig:aloop}} &
        \subfloat[][Quasi-loop]{\includegraphics{fig-6.eps}\label{fig:qloop}}
      \end{tabular}
    \end{tabular}
  \end{tabular}
  \caption{Dashed interaction graph and three examples}
  \label{fig:dashed}
\end{figure}

Before we actually give the proof, let us first examine several simple
examples. The first example considers a chain of interactions as in
Fig.~\ref{fig:achain}. Let $\ket{\alpha_j} \otimes \ket{\beta_j}$ be
the constraint on the $j$-th edge. We will call it an alternating
chain if $\ket{\beta_{j-1}}$ and $\ket{\alpha_j}$ are linearly
independent for all $j$. It is easy to see that the solution space is
$k+1$ for an alternating chain of $k$ qubits. The second example shown
in Fig.~\ref{fig:aloop} is called the alternating loop. As its name
suggests, it is a loop where the two constraints on any vertex are
linearly independent. Any alternating loop has solution space of
dimension $2$, namely the span of \ket{00\ldots 0} and \ket{11\ldots
  1} up to the local operation that maps $\ket{\alpha_j}$ and
$\ket{\beta_{j-1}}$ to $\ket{0}$ and $\ket{1}$. The final example we
consider is called the quasi-alternating loop. It is almost the same
as the alternating loop except that there is one special vertex on the
loop having the same constraint on the two edges adjacent to it.
Figure~\ref{fig:qloop} gives such an example where the top vertex is
special. It is easy to see that the constraint on the special vertex
of a quasi-alternating loop must be satisfied. In particular, for the
loop in Fig.~\ref{fig:qloop}, the top vertex must be \ket{1} as
otherwise it will be impossible to satisfy all five constraints on the
loop.

We now start the proof by induction on $n$, the number of qubits. For
$n=1,2$, the observation is trivial. If there is a vertex $a$ on which
the constraints are the same up to global phases, let the constraints
be $\ket{0}_a$ and, more concretely, let the constraints on an edge
connects to $a$ be of the form $\ket{0}_a\ket{\alpha}_b$ for some
qubit $b$. We can write any state in the ground space as $\ket{\Phi} =
\ket{0}_a\ket{\Phi_0} + \ket{1}_a\ket{\Phi_1}$. Obviously,
$\ket{\Phi_0}$ and $\ket{\Phi_1}$ are both in the ground space of the
constraints not acting on $a$. Moreover, $\ket{\Phi_0}$ also needs to
be orthogonal to $\ket{\alpha}_b$'s. By the induction hypothesis, both
\ket{\Phi_0} and \ket{\Phi_1} are in a product span. Therefore,
$\ket{\Phi}$ is also in a product span. On the other hand, if one
cannot find any vertex whose constraints are the same, we can find
either an alternating loop or a quasi-alternating loop in the graph.
If a quasi-alternating loop is found, we know the state for the
special vertex of the loop and can use the induction hypothesis on the
remaining system. Otherwise, if an alternating loop is found, we can
write any state in the ground space as
\begin{equation}
  \ket{\Phi} = \ket{00\cdots 0}\ket{\Psi_0} + \ket{11\cdots 1}\ket{\Psi_1},
\end{equation}
up to local operations on the loop. If a constraint acts on two qubits
on the loop, it can only restricts the loop to be exactly
$\ket{00\cdots 0}$ or $\ket{11\cdots 1}$. The analysis is similar to
the first case when a constraint $\ket{\alpha}_a\ket{\beta}_b$ acts on
one qubit $a$ on the loop and another qubit $b$ outside of the loop.
This completes the proof. Notice that the local operations chosen here
are determined by the constraints of alternating loops, and that one
will never have two alternating loops giving different local
operations for a single qubit, the orthogonality of the states up to
local operations follows. We note that the orthogonality property only
holds for the product constraints. The symmetric subspace, for
example, is not a span of {\it orthogonal} product states up to local
operations although it is the span of $\ket{00}, \ket{11},
\ket{{+}{+}}$ where $\ket{+} = (\ket{0} + \ket{1}) / \sqrt{2}$.

\begin{figure}[htbp]
  \centering
  \begin{tabular}{m{.35\linewidth}m{.35\linewidth}}
    \begin{tabular}{c}
      \subfloat[][Type-I Sliding]{\includegraphics{fig-3.eps}
        \label{fig:sliding_1}}\\
      \subfloat[][Type-II Sliding]{\includegraphics{fig-4.eps}
        \label{fig:sliding_2}}
    \end{tabular}
    & \subfloat[][An example of simplified interaction graph]{
      \includegraphics{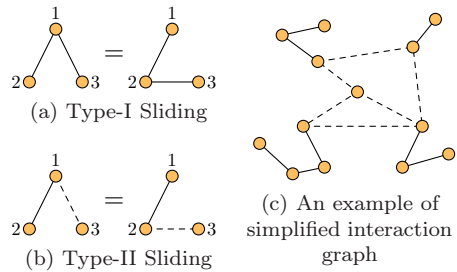}\label{fig:toy_solid}}
  \end{tabular}
  \caption{Simplification of the interaction graph}
  \label{fig:solid}
\end{figure}

{\em General homogeneous case.---\/} Given a general homogeneous
Hamiltonian, the interaction graph will consist both solid and dashed
edges. The main technique is to simplify the interaction graph in hand
without changing the ground space. Two sliding operations as shown in
Figs.~\ref{fig:sliding_1}~and~\ref{fig:sliding_2} will be used in the
simplification. The {\tt Type-I} sliding says that if we have
entangled interactions between $1,2$ and $1,3$, we can change it to
two entangled interactions between $1,2$ and $2,3$ without affecting
the ground space. The {\tt Type-II} sliding is of a similar spirit,
but involves both entangled and product interactions. We will only
prove the validity of {\tt Type-I} sliding as a similar argument holds
for the {\tt Type-II} sliding operation.

Let $\Pi_{12}=\oprod{\phi}{\phi}$ and $\Pi_{13}=\oprod{\psi}{\psi}$ be
the two rank-$1$ operators acting on qubit $1,2$ and $1,3$. We will
find a local interaction $\Pi_{23}$ acting on $2,3$ such that
$\Pi_{12}+\Pi_{23}$ has the same ground space as $\Pi_{12}+\Pi_{13}$.
As $\ket{\phi}$ and $\ket{\psi}$ are entangled states, one can find
local operations $L_2$ and $L_3$ acting on qubit $2$ and $3$
respectively such that $\ket{\phi} = I_1\otimes L_2\ket{Y}$ and
$\ket{\psi} = I_1 \otimes L_3\ket{Y}$ where $\ket{Y}$ is the
singlet state $(\ket{01}-\ket{10})/\sqrt{2}$. The ground space of
$\Pi_{12}+\Pi_{13}$ is therefore
\begin{equation*}
\begin{split}
  & \mathcal{K}(I\otimes L_2\oprod{Y}{Y}_{12} I\otimes L_2^\dagger + 
  I\otimes L_3\oprod{Y}{Y}_{13} I\otimes L_3^\dagger)\\
  =\; & (L^\dagger)^{-1} \mathcal{K}(\oprod{Y}{Y}_{12}+\oprod{Y}{Y}_{13})\\
  =\; & (L^\dagger)^{-1} \mathcal{K}(\oprod{Y}{Y}_{12}+\oprod{Y}{Y}_{23})\\
  =\; & \mathcal{K}(\Pi_{12} + L_2\otimes L_3 \oprod{Y}{Y}_{23} 
  L_2^\dagger\otimes L_3^\dagger),
\end{split}
\end{equation*}
where the first equation uses Eq.~\eqref{eq:L}, the second one is
obtained by a direct calculation establishing that
$\mathcal{K}(\oprod{Y}{Y}_{12}+\oprod{Y}{Y}_{13})$ is the symmetric
subspace of the three qubits, and the last step employs
Eq.~\eqref{eq:L} again. This validates the {\tt Type-I} sliding
operation.

Repeated applications of the two types of sliding operations can
modify an arbitrary graph (a homogeneous Hamiltonian) with solid and
dashed edges to the so-called simplified interaction graph (simplified
homogeneous Hamiltonian). The simplified graph has a backbone of only
dashed edges and several solid-edge tails attached to the backbone. An
example of such a graph is shown in Fig.~\ref{fig:toy_solid}. This
simplification can be done in two steps by first changing each
connected component of solid edges into a tail, and then sliding all
dashed edges connected to a tail to one end of the tail. During the
process of the sliding operations, it may happen that there is more
than one edge between two vertices. If these multiple edges represent
different constraints, one will essentially have a high rank
constraint and can deal with it as before in the case study of rank.

{\em Simplified homogeneous case.---\/}
Since sliding operations do not change the ground space, we only need
to work with simplified interaction graphs. The idea is to build the
entire ground space by extending the ground space for the dashed
backbone. Let us first consider the case where there is only one tail
in the simplified interaction graph. More specifically, let $S$ be the
ground space of the dashed constraints in the backbone $J$, and $T$ be
the symmetric subspace confined by the tail of qubit set $K$, where
$J\cap K$ has exactly one qubit $a$, through which the tail is
attached to the backbone. We prove that $R = S \otimes \mathcal{H}_{K
  \setminus \{a\}} \cap T \otimes \mathcal{H}_{J \setminus \{a\}}$ is
again a product span. Write $S$ as the direct sum
\begin{equation*}
  \left( S_0 \otimes \mathcal{H}_a \right)\oplus
  \left(\bigoplus_{j=1}^d S_j \otimes \ket{\alpha_j^\perp}_a \right),
\end{equation*}
where $\ket{\alpha_j}_a$'s are different dashed constraints on vertex
$a$ and $d$ is number of such $\ket{\alpha_j}_a$'s. For the basis of
$S_0$, all the constraints in the backbone are already satisfied, and
therefore, the qubit $a$ can be any state. We say that qubit $a$ is
free in this case. For the basis of $S_j$, qubit $a$ has to be
$\ket{\alpha_j^\perp}$ in order to satisfy all the constraints in the
backbone. In this case, the state can only be extended to the tail by
copying. In summary, the intersection $R$ contains the space
\begin{equation}\label{eq:R}
  \left( S_0 \otimes T \right)\oplus
  \left(
    \bigoplus_{j=1}^d S_j \otimes \ket{\alpha_j^\perp}^{\otimes |T|}
  \right).
\end{equation}
We will need to show that this is actually everything in $R$. 

We first claim that the product basis for $S_j$'s all together form a
linearly independent set. By orthogonality (up to local operations),
$S_j$ and $S_k$ are orthogonal if $\ket{\alpha_j}$ and
$\ket{\alpha_k}$ are not. On the other hand, if $\ket{\alpha_j}$ and
$\ket{\alpha_k}$ are orthogonal, the basis for $S_j$ and $S_k$ are
linearly independent. Otherwise, we will find a state $\ket{\psi}$ in
both $S_j$ and $S_k$, meaning that $\ket{\psi}$ should be in $S_0$, a
contradiction. Now, for any state $\ket{\Psi}$ in $R$, we can write it
as $\ket{\Psi} = \sum_j \ket{\Psi_j} \ket{\Phi_j}$ where
$\ket{\Psi_j}$'s are linearly independent product states spanning $S$.
Let $\ket{\hat{\Psi}_j}$ be the state on $J\setminus \{a\}$ when the
state on $J$ is $\ket{\Psi_j}$. One can also collect terms according
to the state on $J\setminus\{a\}$, that is, $\ket{\Psi} = \sum_k
\ket{\hat{\Psi}_k} \sum_l \ket{\Psi}^a_{k,l} \ket{\Phi_{k,l}}$. As
shown previously, $\ket{\hat{\Psi}_k}$'s are linearly independent, and
we know $\sum_l \ket{\Psi}^a_{k,l} \ket{\Phi_{k,l}}$ is in $T$ for
each $k$. That is, the state $\ket{\psi}$ is indeed in the space of
Eq.~\eqref{eq:R}. As the symmetric subspace can always be spanned by
product states, we have finished the proof for the case of one tail. For
multiple tails, the proof is essentially the same by an induction on the
number of tails.

{\em Application to the counting of degeneracy.---\/} The results
above actually allow us to prove that counting the ground state
degeneracy of a 2BFF Hamiltonian is in \#P. The class \#P contains
functions $f$ if there is a polynomial time algorithm $A$ such that
\begin{equation*}
  f(x) = |\{y,A(x,y)\text{ accepts.}\}|,
\end{equation*}
where $y$ is usually called a proof to the verifier $A$.

As indicated by the ground space structure in Fig.~\ref{fig:forest},
the isometries will not change the dimension and we only need to
consider the simplified homogeneous case where one can actually
replace the solid edges of the tails to be dashed edges forming
alternating chains. As long as we choose the constraint of the tail on
the vertex connecting to the backbone to be different from all other
constraints $\ket{\alpha_j}$ of that vertex, the dimension of the
solution space remains unchanged. To understand this, we need to
review the extension of the product span with intersection of
symmetric subspaces. If the vertex in the intersection is free, we
will have the whole symmetric subspace on the tail which is of
dimension $k+1$ where $k$ the number of qubits in the tail. This
coincides with the dimension of the alternating chain. If the vertex
in the intersection is not free, we will have a unique extension in
the tail, which again coincides with the case of alternating chain.

It therefore suffices to count the dimension of any dashed graph. To
show that it is in \#P, one can choose the proof to the verifier to be
the non-deterministic $0,1$ choices in the case of (1)
all-the-same-constraint vertex and (2) alternating loop.


\begin{acknowledgments}
  We thank S. Bravyi and X.G. Wen for valuable discussions. ZJ
  acknowledges support from NSF of China (Grant Nos. 60736011 and
  60721061); his research at Perimeter Institute is supported by the
  Government of Canada through Industry Canada and by the Province of
  Ontario through the Ministry of Research \& Innovation. ZW is
  supported by the grant from the Centre for Quantum Technology, the
  WBS grant under contract no. R-710-000-008-271. BZ is supported by
  NSERC and CIFAR.
\end{acknowledgments}

\bibliography{structure}


\end{document}